\documentstyle[preprint,prd,aps,psfig,floats,tighten]{revtex}

\begin{document}

\draft
\preprint{OKHEP--97--06}

\title{Analytic perturbation theory: A new approach to the
analytic continuation of the strong coupling constant $\alpha_S$
into the timelike region}

\author{Kimball A. Milton\thanks{E-mail: milton@mail.nhn.ou.edu}}
\address{Department of Physics and Astronomy, The University of
Oklahoma,
Norman, OK 73019 USA}
\author{Olga P. Solovtsova\thanks{E-mail:
olsol@thsun1.jinr.dubna.su}}
\address{Bogoliubov Laboratory of
Theoretical Physics, Joint Institute for Nuclear Research\\
Dubna, Moscow Region, 141980, Russia}
\date{\today}
\maketitle

\begin{abstract}
The renormalization group applied to perturbation theory is
ordinarily
used to define the running coupling constant in the spacelike region.
However, to describe processes with timelike momenta transfers,
it is important to have a self-consistent determination of the
running coupling constant in the timelike region.
The technique called analytic perturbation theory (APT)
allows a consistent determination of this running coupling constant.
The results are found to disagree significantly with those obtained
in the standard perturbative approach.  Comparison between the standard
approach and APT is carried out to two loops, and threshold matching in
APT is applied in the timelike region.
\end{abstract}
\pacs{11.10.Hi,11.55.Fv,13.35.Dx,12.38.Cy}

\section{Introduction}
A fundamental issue in quantum chromodynamics (QCD) is
the behavior of the strong interaction running coupling constant
 $\alpha_S=g_S^2/4\pi$.
The basic research tool is perturbation theory (PT)
with its renormalization-group improvement~\cite{BSh}.
In the  QCD case in the limit of large momentum transfers $Q$,
this approach provides a logarithmic decrease of the running coupling
constant $\alpha_S \sim 1/\ln(Q^2/\Lambda^2)$, where
$\Lambda$ is the QCD scale parameter that determines
where the theory becomes asymptotically free.
The study of the behavior of $\alpha_S$ outside of the asymptotic
region is more difficult.  It is known that the
direct use of PT improved by the renormalization group
leads to infrared instability of $\alpha_S$ and unphysical
singularities, for instance, a ghost pole at $Q^2=\Lambda^2$.
Unphysical singularities of a perturbative running coupling constant
precludes a self-consistent determination of the effective coupling
constant for timelike momentum transfers.
Recently, a new method has been proposed \cite{DVS} for constructing
the QCD running coupling constant in such a way as to retain the
correct analytic properties. This method is called analytic
perturbation theory (APT). The main purpose of this paper is to
analyze the region of timelike momentum transfers on the basis of APT
\cite{DVS,KS1}, and compare the results of the PT and APT approaches.

It is well known that a theoretical description of important timelike
processes such as $e^+e^-$ annihilation into hadrons, or of decay
widths of the $\tau$-lepton and $Z$-boson into hadrons, requires
analytic continuation of the running coupling constant from
the spacelike (Euclidean) region of momentum transfers
($q^2=-Q^2<0$) into the timelike (physical) region ($q^2>0$).
Although this problem has been studied since the 1970's
(see, e.g., \cite{MPS}), it still remains a subject of
great interest (see, e.g., \cite{SurS,Raczka,Rad}).
It is obvious that information on the running coupling constant
obtained from timelike processes, for instance,
from $e^+e^-$ annihilation into hadrons, corresponds to knowledge of
the coupling constant extracted from spacelike processes such as deep
inelastic scattering, if the transition from the Euclidean into
the physical region is performed in a correct manner (see~\cite{KS1,JS1})
without violation of analytic properties of the hadronic correlator
$\Pi(q^2)$ and the Adler function $D(q^2)$.
When the analytic properties are not respected,
the question arises: To what extent does this breaking
of analyticity influence quantities extracted from physical
processes? It is impossible to answer this question within the framework of
standard perturbation theory. On the other hand,
the APT method retains the correct analytic properties of the Adler
$D$-function and, in addition, gives simple analytic expressions that can be
compared with corresponding PT expressions and, therefore, allows quantitative
analysis of the influence that the breaking of $Q^2$-analyticity has on
the running coupling constant.

The organization of this paper is as follows:  In the following section,
we discuss the procedure of analytic continuation from the spacelike
(Euclidean) to the timelike ($s$-channel) region.  In Sec.~III
we examine this procedure in the conventional one-loop PT approach, and
demonstrate that it is inconsistent.  In Sec.~IV we
resolve this problem through the APT approach, and in Sec.~V we
compare the results of these two schemes.  We move on to two-loops
in Sec.~VI, and demonstrate the stability of the APT approach.
A matching procedure for timelike momentum transfers is given in
Sec.~VII, where we show how the coupling constant depends on the
number of active flavors.  A summary of our results is given in the
Conclusions.

\section{Effective coupling constants in the timelike and spacelike
regions}

First, we note that in the standard approach,
the running coupling constant in QCD as a function of $Q^2$
is determined by the renormalization-group analysis in the
region of spacelike momentum transfers.
However, to parametrize many physical processes, one needs to know
the coupling constant in the timelike region. To be specific, many
experimentally measured ratios $R_{\sigma}$,
where, e.g.,  $\sigma = e^+e^-,\; \tau \;, \; Z \ldots \;$,
can be written in the form
$R_{\sigma}  = R_{\sigma}^{(0)}( 1 + \Delta_{\sigma} )$.
Here $\Delta_{\sigma}\,$ is a QCD correction and  $R_{\sigma}^{(0)}$
represents the parton level of description of a given process with
electroweak corrections.
$\Delta_{\sigma}$  can be expressed through the imaginary part of the
hadronic correlator ${\cal R}(s)={\rm Im}\; \Pi(s)/\pi$.
To parametrize ${\cal R}(s)$ in terms of QCD parameters, a special
procedure of analytic continuation is required.
With that end in view, one usually employs a dispersion relation
\begin{equation}
\label{D2}
D(z)=-z\frac{d \Pi(z)}{dz}=
-\,z\int_0^{\infty}\,ds
\frac{{\cal R}(s)}{{(s-z)}^2} \, ,
\end{equation}
where $z=q^2$, and the inverse relation
\begin{equation}
\label{R2}
{\cal R}(s) =
\frac{1}{2\pi {\rm i}}\,
\int _{s-{\rm i}\,\epsilon} ^{s+{\rm i}\,\epsilon} dz
\frac{d \Pi(z)}{dz}=
-\frac{1}{2\pi {\rm i } }  \int _{s-{\rm i}\,
\epsilon} ^{s+{\rm i}\,\epsilon} dz
\frac{D(z)}{z} \, ,
\end{equation}
where the contour joins points $s\mp{\rm i}\,\epsilon$
and lies in the region of analyticity of the function  $D(z)$,
going around the cut ${\rm Re}\; z>0$.

We define the effective  coupling constants $\bar{a}^{{\rm eff}}$
and $\bar{a}^{{\rm eff}}_s$,
respectively, in the spacelike
($t$-channel) and timelike ($s$-channel)
regions, using the notation ${\bar a}={\bar \alpha}/4\pi$
and dimensionless
(in units of the scaling parameter $\Lambda$)  momentum variables, by
\begin{eqnarray}
\label{D}
& &D(z) \,\propto\, 1\,+\,d_1\,{\bar a}(z)\,+\,d_2\,{\bar a}^2(z)\,
+ \cdots \,
=1+d_1\,\bar{a}^{{\rm eff}}(z)\, ,
\\
\label{R}
& &{\cal R}(s) \,\propto\, 1\,+\,r_1\,{\bar a}_s(s)\,+\,r_2\,{\bar a}_s^2(s)\,
+\cdots =1+r_1\,{\bar{a}}^{{\rm eff}}_s(s)\, .
\end{eqnarray}
Relations (\ref{D2}) and (\ref{R2}) and the equality $d_1=r_1$
result in the connection between these effective coupling constants
\begin{eqnarray}
\label{as0}
& & \bar{a}^{{\rm eff}}_s(s)=
-\,\frac{1}{2\pi {\rm i}}\,
\int _{s-{\rm i}\,\epsilon} ^{s+{\rm i}\,\epsilon} \frac{dz}{z}
\,\bar{a}^{{\rm eff}}(z) \,  , \\  \label{at0}
& &\bar{a}^{{\rm eff}}(z)=-\,z\,\int_0^{\infty}\,
\frac{ds}{{(s-z)}^2}\,\, \bar{a}^{{\rm eff}}_s(s)\, .
\end{eqnarray}
Therefore, the QCD corrections $\Delta_{\sigma}\,$ for the class of physical
processes considered with timelike momentum transfers are to be parametrized,
according to Eq.~(\ref{R}), by the effective coupling constant
$\bar{a}^{{\rm eff}}_s(s)$, which is explicitly related to
$\bar{a}^{{\rm eff}}(z)$ by Eqs.~(\ref{as0}) and (\ref{at0}).

In any finite order of PT, the analytic properties
of the running coupling $\bar{a}(z)$ should be the same as for
the effective coupling constant $\bar{a}^{{\rm eff}}(z)$.
Therefore, the connection between $t$- and $s$-channel running coupling
constants, $\bar{a}(z)$ and $\bar{a}_s(s)$ is defined by equations like
(\ref{as0}) and (\ref{at0}).
In the one-loop approximation the effective coupling
constants coincide with the running coupling constants 
and in higher loops, the connection depends on the physical process.

\section{PT analysis}

Consider the above procedure of analytic continuation within PT. In
the one-loop approximation, the running coupling constant is of the form
\begin{equation}
\bar{a}^{{\rm PT}}(z)\,=\,
\frac{1}{\beta_0}\,\frac{1}{\ln\,(Q^2/\Lambda^2)}
\,=\,  \frac{1}{\beta_0}\,\frac{1}{\ln\,(-z)} \,
,\quad z \equiv -Q^2/\Lambda^2 \> ,
\label{a1z}
\end{equation}
where $\beta_0=11-2n_f/3$  is the one-loop coefficient of the
$\beta$-function
corresponding to $n_f$ active quarks.
Inserting Eq.~(\ref{a1z}) into Eq.~(\ref{as0}) we obtain the
following expression for the running coupling constant in $s$-channel
\begin{equation}
\label{raz1}
 \bar{a}_s^{{\rm PT}}(s)=-\frac{1}{2\pi {\rm i}\beta_0}
\int _{s-{\rm i}\,\epsilon} ^{s+{\rm i}\,\epsilon}
\frac{d z}{z}\; \frac{1}{\ln (-z)} =
-\frac{1}{\pi\beta_0} \left( \frac{\pi}{2}+
\arctan\frac{\ln s}{\pi}   \right) \, .
\end{equation}
This expression is physically meaningless, because it is negative for
any $s$ and does not have the correct asymptotics, that is, going as
$1/\ln s$ as $s \to \infty$; the reason will be explained below.

\begin{figure}[hpt]
\centerline{ \psfig{file=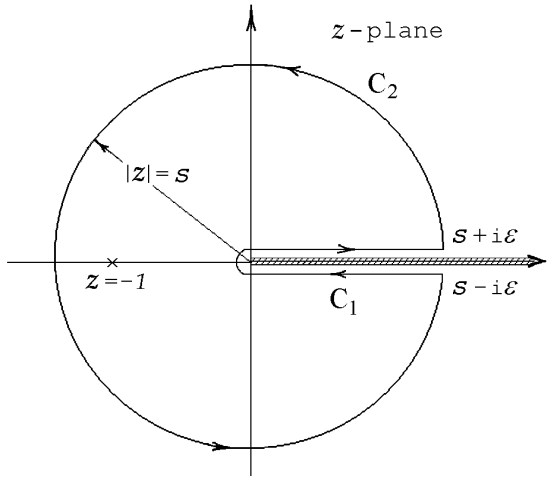,width=8.0cm}}
\caption{\sl
 Integration contours in the complex $z$-plane. }
       	\label{circle}
         \end{figure}

There is another way of calculating $\bar{a}_s$, based on the Shankar
method~\cite{Shankar}.
Using analyticity of the $D$-function in the complex $z$-plane
with the cut along the positive real
axis, we may pass from the integral along the cut,
given by expression (\ref{as0}),
that is, around the contour $C_1$ (see Fig.~\ref{circle}),
to an integral around a circle of radius $|z|=s$ in the
complex $z$-plane, contour $C_2$, parametrized by
$ z=-s \exp({\rm i} \varphi)$,
$-\pi < \varphi <  \pi $, to arrive the expression
\begin{equation}
\label{cont1}
{\bar a}_s^{\rm ``PT"}(s)=\frac{1}{2\pi {\rm i}}\,
\int_{C_2} \frac{dz}{z}\; \bar{a}^{{\rm PT}}(z) =\frac{1}{2\pi\beta_0} \;
\int_{-\pi}^{\pi} \frac{d \varphi}{\ln s + {\rm i} \varphi} =
\frac{1}{\pi\beta_0} \arctan\frac{\pi}{\ln s};
\end{equation}
this is positive when $s>1$ and possesses the correct ultraviolet
asymptotics.
It is just this expression that is used as a one-loop PT result for
all timelike momenta $s>0$:
\begin{equation}
\label{PTs}
{\bar \alpha}_s^{``\rm PT"}(s)= \frac{4}{\beta_0} \arctan
\frac{\pi}{\ln s} \; .
\end{equation}
It is obvious that Eq.~(\ref{PTs}) provides the restriction
${\bar \alpha}_s^{``\rm PT"}(s) \leq 2\pi/\beta_0$ for any $s$
(see \cite{PivNC}).

Thus, a formal conversion of the PT one-loop running coupling
constant in the spacelike region~(\ref{a1z})
into expressions for the coupling constant
in the timelike region leads to contradictory results (\ref{raz1})
and (\ref{cont1}). The reason can easily be understood if
one applies the Cauchy theorem (see Fig.~1)
to establish the connection between the integrals in
Eqs.~(\ref{raz1}) and (\ref{cont1}),
\begin{equation}
\label{Coshi}
\frac{1}{2\pi {\rm i}}\,
\int_{C_2} dz \,\psi(z)=
-\,\frac{1}{2\pi {\rm i}}
\int _{C_1}dz \,\psi(z) + {\rm res}\;\left[\,\psi(z),-1 \right] \; , \, \,
\psi(z)\equiv \frac{1}{  z\ln (-z)} \, ,
\end{equation}
which is consistent with Eq.~(\ref{raz1}) and (\ref{cont1}) because
the residue of the function $\psi(z)$ at the point $z=-1$ is $1$.
Therefore,  the discrepancy between Eq.~(\ref{raz1})
and Eq.~(\ref{cont1}) is due to an unphysical ghost pole in
Eq.~(\ref{a1z})
at $z=-1$ that violates the required analytic properties of the
running coupling constant. The inclusion of multiloop corrections does
not solve this problem but rather
produces new unphysical singularities, as we will see in Sec.~VI.
Therefore, keeping to standard
PT approximations that violate the necessary analytic properties of
the running coupling constant makes it impossible to pass into the
timelike region in a self-consistent way. This can, for instance,
be demonstrated by making
an inverse transition from the timelike into the spacelike region
with the help the dispersion relation~(\ref{D2}).
Substituting the running coupling constant $\bar{a}_s(s)$ given by
Eq.~(\ref{cont1}) into integral (\ref{at0}) following from
Eq.~(\ref{D2})
and taking account of the expression $\arctan( {\pi}/{\ln s})
\,=\, {\rm sgn} (\ln s)\,\pi/2- \arctan(\ln s/\pi)\, $,
we arrive at the formula
\begin{equation}
\label{newa1z}
\bar{a}^{\rm ``PT"}(z)\,=\,
\frac{1}{\beta_0}\,
\left[\, \frac{1}{\ln \left(-z\right)}\,+\,\frac{1}{1+z}
-\,\frac{1}{1-z}
\,\right] \>,
\end{equation}
which is different from the starting point, Eq.~(\ref{a1z}).

\section{APT analysis}
The problem of how to make the
correct transition between the space- and timelike regions
can  be solved in the framework of the APT method \cite{DVS,KS1}
that ensures  the correct analytic properties of the coupling
constant without introducing extra parameters. The resulting one-loop
expression for the analytic coupling constant in the Euclidean region
is as follows:
\begin{equation}
\label{aan1}
{{\bar a}}^{\rm APT}(z)\,=\,
\, \frac{1}{\beta_0}\,
\left[\, \frac{1}{\ln \left(-z\right)}\,+\,\frac{1}{1+z} \right ] \>.
\end{equation}
The first term in brackets determines the asymptotic behavior at
large momenta
and is of the form given in PT. The second term, of a nonperturbative
nature, compensates the ghost pole at $z=-1$.
When one employs the analytic coupling constant (\ref{aan1}),
both methods of calculating $\bar{a}_s$ considered above produce
the same result, i.e.
\begin{equation}
\label{newCoshi}
{\bar a}_s^{\rm APT}(s)\,=\,
-\frac{1}{2\pi {\rm i}}\,\int_{C_1}\frac{dz}{z}{\bar a}^{\rm APT}(z)\,
=\, \frac{1}{2\pi {\rm i}}\,
\int_{C_2} \frac{dz}{z} {\bar a}^{\rm APT}(z)\,
= \frac{1}{\pi\beta_0} \left(\frac{\pi}{2}-
\arctan \frac{\ln s}{\pi} \right).
\label{exact1}
\end{equation}
We note that APT and PT give distinct values for the running coupling
constant in the timelike region, for example,
\begin{equation}
{\bar a}_s^{\rm APT}={\bar a}_s^{\rm PT}+{1\over\beta_0},
\end{equation}
where ${\bar a}_s^{\rm PT}$ is given by Eq.~(\ref{raz1}).
Consistency of the APT approach also follows from
the fact that we can reconstruct the initial expression (\ref{aan1})
when
the timelike coupling (\ref{exact1}) is substituted into
Eq.~(\ref{at0}).
It is of interest to note that this consistency is due to the second
term in Eq.~(\ref{aan1}) that compensates the pole, whose contribution to
the integral around the contour $C_2$
 equals zero when $s > 1$, i.e., we have the equality
\begin{equation}
\label{equ}
\int_{C_2} \frac{dz}{z} {\bar a}^{\rm APT}(z)\, =
\int_{C_2} \frac{dz}{z} {\bar a}^{\rm PT}(z) \,,\>  \> \> \> \>
s > 1 \>,
\end{equation}
where the function ${\bar a}^{\rm PT}(z)$ is defined by Eq.~(\ref{a1z}).
Therefore the PT expression (\ref{a1z}) gives the same result as  the
APT approach in the timelike region for $s >1$ if the contour $C_2$
is used. However, there is no inverse correspondence for PT
[see formula (\ref{newa1z})].
Moreover, note that an equality analogous to Eq.~(\ref{equ})
does not arise if the integrand contains the running
coupling constant multiplied by a function of $z$.
For the $R_{\tau}$-ratio, for instance,
$\bar a$ is multiplied by a polynomial in $z$ and, as
is shown in \cite{Oly}, the contour integral over $C_2$ in PT turns
out to be different from that in the APT
approach.\footnote{A detailed comparison of inclusive $\tau$ decay can
be found in Ref.~\cite{MSS1}}

	       \begin{figure}[hbt]
\centerline{ \psfig{file=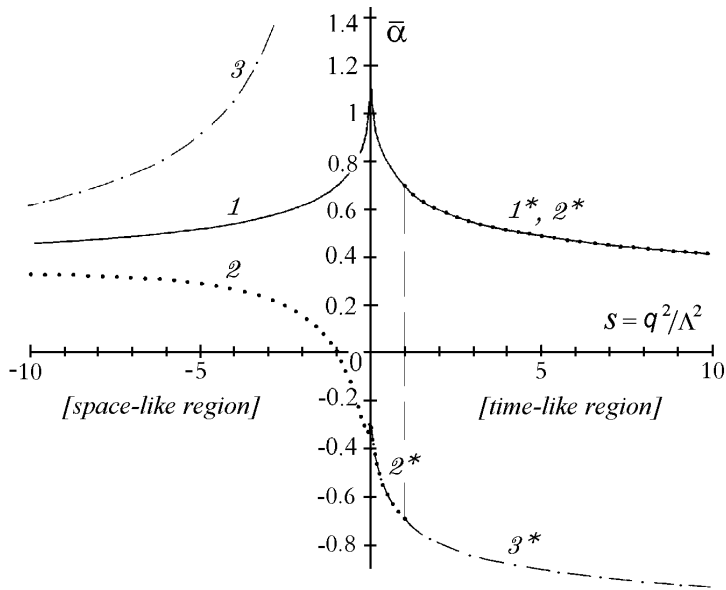,width=10.7cm}}
\caption{ \sl
The behavior of the running coupling constant
calculated by different methods as a
function of the dimensionless variable $s=q^2/\Lambda^2$.
  }
\end{figure}

\section{Numerical comparison of running coupling constants}
The results obtained are illustrated in a series of figures.
(In these figures we take the number of quark flavors to be 3.)
Firstly, we consider the region of small momentum transfers.
Fig.~2 shows the behavior of the running coupling constant
computed by different methods in the region
$-10 \leq s=q^2/\Lambda^2 \leq 10$.
The solid line represents the APT coupling constant calculated by
formula (\ref{aan1}) in the spacelike region
(number $\it 1$ in the figure) and by formula
(\ref{exact1}) in the time-like region (curve ${\it 1^*}$).
Dots denote the ``PT'' coupling constant determined by
Eq.~(\ref{PTs}) (curve ${\it 2^*}$) and by
Eq.~(\ref{newa1z}) (curve ${\it 2}$).
The dash-dotted line ${\it 3}$ represents the PT coupling constant
computed by formula (\ref{a1z}) in the spacelike region,
and curve ${\it 3^*}$ by formula (\ref{raz1}) in the timelike region.
(Incidentally, note that curves {\it 2} and ${\it 2^*}$ vanish at the
origin, which is beyond the resolution of this figure, while curves {\it 1}
and ${\it 1^*}$ approach the universal value $4\pi/\beta_0=1.40$ at
the origin.)

As is seen from Fig.~2, the behavior of the APT
running coupling constants (curves ${\it 1}$ and ${\it 1^*}$) is
almost, but not quite, mirror-symmetric,
and at $s=0$   the space- and timelike APT running coupling constants
are both equal to the universal value $4\pi/\beta_0$
(see \cite{KS1}). The pairs of curves of the standard PT approach,
${\it 3}$ and  ${\it 2^*}$, or ${\it 3}$ and ${\it 3^*}$,
do not show analogous behavior of the running coupling constants.
In the spacelike region the function ${\bar\alpha}^{\rm PT}$
grows without limit (curve ${\it 3}$), whereas in the timelike
region (curve ${\it 2^*}$) it is limited to the value $2\pi/\beta_0$.
Curve ${\it 2}$ calculated with the coupling given by
curve ${\it 2^*}$ in the dispersion integral does not reproduce the
initial curve ${\it 3}$.

Now consider the region where the running coupling constant
${\bar \alpha} \sim 0.3 $, which approximately corresponds to the
mass scale of the $\tau$-lepton, $M_{\tau}=1.78 $ GeV,
${\bar \alpha}(M_{\tau})=0.34 \pm 0.04 $~\cite{War}),
defined in the spacelike region.
It is known that the decay
$\tau \to$  hadrons is important for testing QCD, as it allows
the most accurate determination of the running coupling constant
at comparatively low energies (see the review \cite{Pich}).
	       \begin{figure}[hpt]
\centerline{ \psfig{file=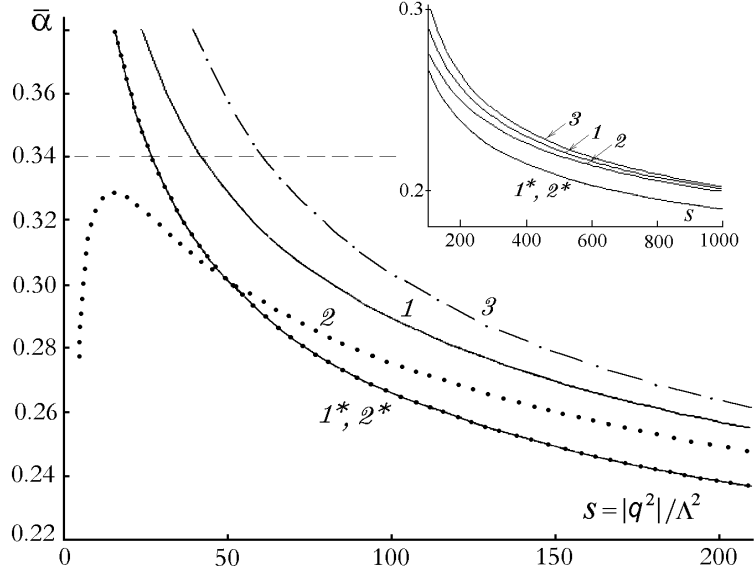,width=10.7cm}}
\caption {\sl
Behavior of the running coupling constant
as a function of the dimensionless variable $ s=|q^2|/\Lambda^2 $.
Notation is the same as in Fig. 2;
a dashed horizontal line corresponds to ${\bar\alpha} =0.34\; $.
The graph on the right top shows the behavior ot the same curves for
large values of $s$.
}
       	\label{f3}
\end{figure}

Fig.~3 shows the behavior of the running coupling constant
versus the dimensionless variable $ s=|q^2|/\Lambda^2 $.
Notation is the same as in Fig. 2;
a dashed horizontal line corresponds to ${\bar\alpha} =0.34\; $.
As it is seen from the Fig.~3, curves ${\it 1,2,3}$
that describe the spacelike region noticeably differ from each other.
With increasing $s$, they begin, as they should, to approach each
other, which is demonstrated on the top right of Fig.~3.
Values of the parameter $\Lambda$
calculated with the running coupling constants described by curves
${\it 1,2}$ and ${\it 3}$ are different. For example, the value of
APT-function (curve ${\it 1}$), equal to  $0.34$ is achieved at
$s_0=41.5$, which corresponds to $\Lambda^{\rm APT}=276$~MeV.
For PT-curve ${\it 3}$ $s_0=60.5$ and $\Lambda^{\rm PT}=228$~MeV.
Note that for curve ${\it 2}$
the value ${\bar\alpha} =0.34$ cannot be achieved at any value of
$s$. For timelike momentum transfers, recall that curves
${\it 1^*}$ and  ${\it 2^*}$
as functions of the dimensionless variable $s$ coincide when $s>1$.
However, they are characterized by different values of $\Lambda$,
which results in different values of
the running coupling constant in the timelike region,
${\bar \alpha}_s^{\rm APT}(M_{\tau})=0.31$
and ${\bar \alpha}_s^{``\rm PT"}(M_{\tau})=0.29$.
(The procedure here is to use the spacelike value of $\bar\alpha$ to
determine $Q^2/\Lambda^2$, and then, with the same numerical value
of $s$, determine ${\bar \alpha}_s$ from Eq.~(\ref{newCoshi}) and
Eq.~(\ref{PTs}),
respectively.) With the accuracy attained at present for experimental
data on the hadronic decay of the $\tau$ \cite{Pich},
this quantitative discrepancy is becoming significant.
	       \begin{figure}[hpt]
\centerline{ \psfig{file=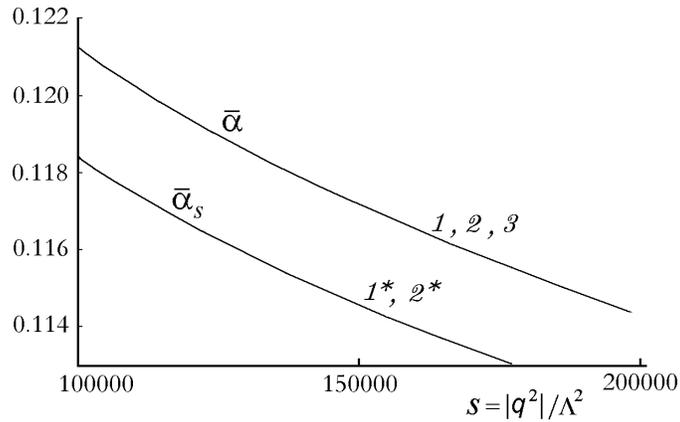,width=9.7cm}}
\caption{\sl
Behavior of the running coupling constant
in the vicinity of the Z-boson mass.
Curves ${\it 1,2,3}$ and ${\it 1^*, 2^*}$ represent the same coupling
constants as in Fig.~2 and Fig.~3.
}
       	\label{f4}
\end{figure}

Let us finally observe the evolution of the running coupling constant
in the region of momentum transfers of order of the $Z$-boson mass
$M_Z=91.2$~GeV. The running coupling constant ${\bar \alpha}$,
corresponding to the dimensionless
variable $s$ is drawn in Fig.~4. (At this point, we neglect the change
in the number of active quarks with growing energy
since this effect does not change the overall picture.  We will consider
the change in the number of active flavors at the two-loop level in Sec.~VII.)
The curve denoted by $\bar\alpha$ represents
all three curves ${\it 1,2}$ and ${\it 3}$ that are drawn in Figs.~2
and 3 and describe the behavior
of the running coupling constant in the spacelike region, which
merge into one curve with  high accuracy for these large values of
$s$. The curve denoted by ${\bar \alpha}_s$ corresponds to the coupling
constant in the timelike region and to curves ${\it 1^*}$ and ${\it 2^*}$
plotted in Figs.~2 and 3.
In the region of sufficiently large values of $s$,
the well-known approximate formula with the so-called $\pi^2$-term
(see, e.g., \cite{Rad,gorishny})
\begin{equation}
\label{pi2}
\bar{\alpha}_s\,=\,{\bar \alpha}\left(
1-\frac{\pi^2}{3}\frac{1}{\ln^2s}
\right) \,
\end{equation}
works well (with an accuracy $\sim 0.1\%$) both for PT and APT.
This approximation gives a difference between
${\bar \alpha}_s$ and ${\bar \alpha}$ of about $ 2 \%$.
Substituting the value of the parameter $\Lambda$ fixed at
$q=M_{\tau}$, we obtain the corresponding values of the
running coupling constant at $q=M_Z$:
\noindent
${\bar \alpha}^{\rm APT}=0.120$, ${\bar \alpha}^{\rm PT}=0.117$
(spacelike region);
${\bar \alpha}_s^{\rm APT}=0.118$, ${\bar \alpha}_s^{\rm PT}=0.114$
(timelike region).
Thus, even at such large values of $s$, the effect of analyticity on
the running coupling constant amounts to $\sim 2\%$,
i.e., it is comparable with the contribution from the $\pi^2$-term
and from higher PT loop corrections.

\section{Two-loop results}

We now extend the above considerations to the two-loop level.
The distinction between the APT and the PT running coupling constants
in the Euclidean region has to do with the unphysical singularities of
the PT running coupling constant.
Following the results of Ref.~\cite{DVS}, we can write down the analytic running
coupling constant in the form of a sum of the standard  perturbative part
and additional terms which compensate for the contributions of the unphysical
singularities, a pole and a cut:
\begin{equation}
\label{a2an}
{\bar a}^{\rm APT}(z)\,=\,  {\bar a}^{\rm PT}(z)\,+\,
\Delta{\bar a}_{\rm pole}(z)\, +\,
\Delta{\bar a}_{\rm cut}(z)\,.
\end{equation}
For the two-loop perturbative running coupling constant,
we use \cite{DVS,MSS1}
\begin{equation}
\label{pt2}
{\bar a}^{\rm PT}(z)\,=\,\frac{1}{\beta_0}\frac{1}{ {\rm L} +
B_1 \ln (1+ {\rm L}/B_1)}~,\quad{\rm L}= \ln(-z)
=\ln\frac{Q^2}{\Lambda^2} \, ,
\end{equation}
where $B_1=\beta_1/{\beta_0}^2$, and $\beta_1=102-38n_f/3$ is the two-loop
coefficient of the $\beta$-function.
Obviously, at large ${\rm L}$ Eq.~(\ref{pt2})
gives the standard PT expression as an expansion in inverse powers of
${\rm L}$,
\begin{equation}
\label{pt_asimpt}
\bar{\alpha}_{\rm asympt}^{\rm PT}
=\frac{4\pi}{\beta_0 {\rm L}}\left(1-B_1\frac{\ln{\rm L}}{\rm L} \right)
+ O \left( \frac{1}{{\rm L}^3} \right) \,.
\end{equation}


According to Eq.~(\ref{pt2}), the contribution coming
from the unphysical pole is cancelled by
\begin{equation}
\label{pole2}
\Delta {\bar a}_{\rm pole}(z)\,=\,
\frac{1}{2 \beta_0}\,\frac{1}{1+z}\, ,
\end{equation}
while the unphysical cut is removed by the following compensation term
\begin{equation}
\label{cut}
\Delta {\bar a}_{\rm cut}(z)\,=\,
\frac{1}{\pi \beta_0}\int_0^{\exp(-B_1)}\,\frac{d\,\sigma}{\sigma\,+\,z}\,
\frac{\pi\,B_1}{\left[\ln{\sigma}\,+\,B_1\,\ln(-1-\ln{\sigma}/B_1)
\right]^2\,+\,\pi^2\,B_1^2}\, .
\end{equation}

To calculate the analytic running coupling constant in timelike region,
we use the expression in terms of the
 spectral density~$\varrho(\sigma)$~\cite{KS1}
\begin{equation}
\label{asrho}
{\bar a}_s^{\rm APT}(s) =\frac{1}{\pi }\,
\int_s^\infty\,\frac{d\sigma}{\sigma}\,\varrho(\sigma)\, ,
\end{equation}
where
$\varrho(\sigma)={\rm Im}\,{\bar a}^{\rm PT}(-\sigma-{\rm i}\epsilon)$.
The spectral density~$\varrho$ plays a central role in the APT method; the
spacelike running coupling constant, ${\bar a}^{\rm APT}$,
is also expressed through~$\varrho$ as follows:
\begin{equation}
\label{arho}
{\bar a}^{\rm APT}(z) =\frac{1}{\pi}\,
\int_0^\infty\,\frac{d\sigma}{\sigma\,-\,z\,-\,{\rm i}\epsilon} \,
\varrho(\sigma) \, .
\end{equation}

As outlined above, independently of the order of approximation,
the APT running coupling constants defined in the space- and timelike
regions at $Q^2=0$ and $s=0$ are both equal to the universal infrared limiting
value $4\pi/\beta_0$~\cite{KS1}, which is important to establish the
stability in the region of the small momentum transfers. (This result is
proved in \cite{MSS1}.) Consider the region
in which the value of running coupling constant ${\bar \alpha} \sim 0.35\, $
(the $\tau$ lepton scale).
Fig.~5 shows the  behavior of the two-loop running coupling constant
for the same interval of the dimensionless variable $s$ as in Fig.~3.
The solid line represents the APT coupling constant in the spacelike
region, which can be computed from (\ref{a2an}); this is like {\it 1} in
Figs.~2 and 3. The dotted line denotes the APT coupling constant in the
timelike region, computed from (\ref{asrho}); this is like {\it 1}$^*$
in one-loop. The dashed line represents the spacelike PT coupling constant
defined by formula~(\ref{pt2}), which is like {\it 3}. The dash-dotted curve
corresponds to timelike PT coupling constant constructed taking into account
$\pi^2$-terms, like Eq.~(\ref{pi2}), and analogous to ${\it 2^*}$.
This figure demonstrates that, as in the one-loop case,
there is a difference in behavior of all these constants.
Moreover, the region in which the value of running coupling constant
is about $0.34$, is shifted to smaller $s$; the value of 0.34 for the
APT running coupling constant is achieved at $s_0=8.6$
which corresponds to $\Lambda^{\rm APT}=607$~MeV and
the value of the timelike running coupling constant
${\bar \alpha}_s^{\rm APT}=0.32$. For the PT running coupling constant,
$s_0=18$ and $\Lambda^{\rm PT}=419$~MeV.
Thus, while in the one-loop case APT parameter $\Lambda$,
is 20\% larger than PT value, in the two-loop case this discrepancy
increases to $45\%$.\footnote{
In Fig.~5 we do not plot the three-loop result~(see \cite{DVS}),
because it practically coincides with the two-loop one.}

	       \begin{figure}[hpt]
\centerline{ \psfig{file=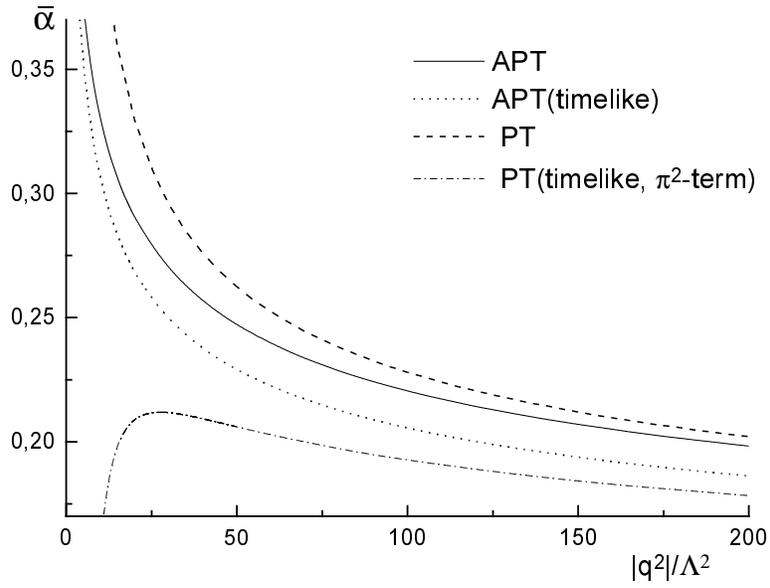,width=13.5cm}}
\caption{\sl
Behavior of the 2-loop running coupling constant
in the PT and APT schemes, both in terms of spacelike and timelike definitions.
}
       	\label{f5}
\end{figure}

In Fig.~6 we show the stability of APT result for
the ratio of the space- and timelike APT running coupling constants.
The solid line corresponds to three-loop case,
the dashed line is two-loop, and the dot-dash is one-loop.
(The one-loop result was already given in \cite{KS1}.)

	       \begin{figure}[hpt]
\centerline{ \psfig{file=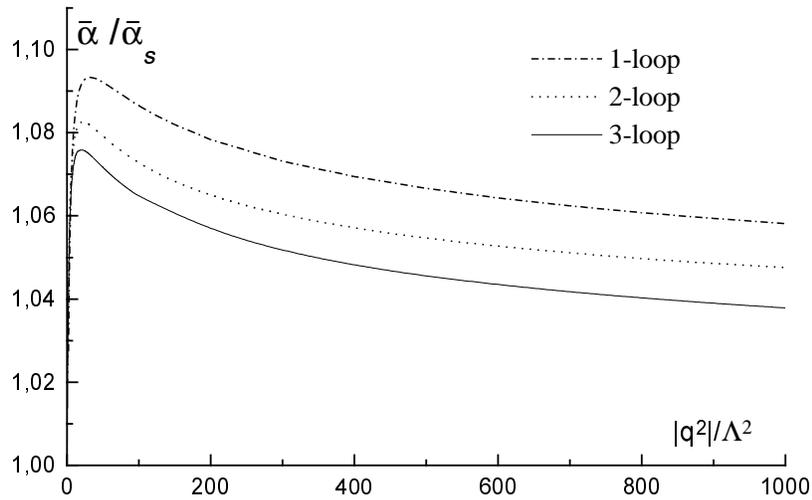,width=13.5cm}}
\caption{\sl The ratio of spacelike and timelike values of the running
coupling constant in the APT scheme, at one, two, and three loops.
}
       	\label{f6}
\end{figure}

Let us now to demonstrate that in the two-loop approximation,
as in the one-loop case, the equality (\ref{equ}) is valid as well.
By using Eq.~(\ref{a2an}), we get
\begin{equation}
\label{equ2}
2\pi {\rm i}\;{\bar a}_s^{\rm APT}=
\int_{C_2} \frac{dz}{z} {\bar a}^{\rm APT}(z)\, =
\int_{C_2} \frac{dz}{z} {\bar a}^{\rm PT}(z) \,
+ \,\int_{C_2} \frac{dz}{z} \Delta {\bar a}_{\rm pole}(z) \,+
 \,\int_{C_2} \frac{dz}{z} \Delta {\bar a}_{\rm cut}(z) \,.
\end{equation}
As follows from~(\ref{pole2}), the pole term
has the same structure as in the one-loop approximation,
and its contribution to contour integral equals zero when $s>1$.
One can also find that the contribution of the cut term~(\ref{cut}) in
Eq.~(\ref{equ2}) equals zero when $s > \exp(-B_1)$.\footnote{
For $0 \leq n_f \leq 6$ we have $0.4 < \exp( -B_1) < 0.6$.   }
As a result, the equality (\ref{equ}) holds for $s>1$ as well, but
it should be stressed (see above discussion in Sec.~V) that this not mean that
the APT and PT timelike coupling constants coincide with each other.

\section{$\lowercase{s}$-channel matching}

Let us now discuss the issue of how the
parameter $\Lambda$ changes with energy as the number of active quark changes:
$\Lambda \to \Lambda_{n_f}\,$ (see \cite{PDG96} for further details.)
The relationship between  $\Lambda_{n_f}$ and $\Lambda_{n_f+1}$
may be fixed by the matching
conditions for coupling constant at
``quark thresholds"~(see, e.g., \cite{MC})
which is usually applied to the
running coupling constant in the Euclidean region.
The APT method opens the new possibility of performing the threshold
matching in the physical region, where the number of active quarks
can be associated with the energy threshold of quark pair production.
It is important to note that any matching procedure of the coupling constant
in the Euclidean region, for which one uses the condition of the type
$ {\rm Re}\,Q^2 > \xi^2\,M_q^2 $ (usually
$ 1 \leq \xi \leq 2$),
leads to a violation of analyticity of $ {\bar a}(z)$.
Although $s$-channel matching demands only continuity of the function
$\bar a_s(s)$ at the threshold, and not of its derivatives, the 
spacelike running coupling constant $\bar a(z)$ will be an analytic function
of $z$, and due to the representation~(\ref{at0})
${\bar a}(z)$ ``knows", in principle, about all quark thresholds.
Therefore, the APT method gives a more consistent definition of
the running coupling constant
and a natural way to perform the matching procedure.
Within the APT approach, we will require that the timelike function
${\bar a}^{\rm APT}_s(s)$ should be a continuous function at the
threshold points:
\begin{equation}
\label{match}
{\bar a}^{\rm APT}_s[(Q_{n_f+1}/\Lambda_{n_f})^2,n_f]
={\bar a}^{\rm APT}_s[(Q_{n_f+1}/\Lambda_{n_f+1})^2,n_f+1] \, ,
\end{equation}
where $Q_{n_f}$ is defined by the pole masses $M_q$ of quark pair.
Taking into account Eqs.~(\ref{equ}) and the results of the previous section
that the unphysical singularities do not contribute at $s>1$ to the contour
integral we  can rewrite Eq.~(\ref{match}) in the following form
\begin{equation}
\label{match-contour}
\int\limits_{|z|=(Q_{n_f+1}/\Lambda_{n_f})^2}\, \frac{dz}{z}
{\bar a}^{\rm PT}(z,n_f)=
\int\limits_{|z|=(Q_{n_f+1}/\Lambda_{n_f+1})^2}\, \frac{dz}{z}
{\bar a}^{\rm PT}(z,n_f+1)\, .
\end{equation}
Therefore, the conventional matching condition
\begin{equation}
\label{pt-match}
{\bar a}^{\rm PT}[(Q_{n_f+1}/\Lambda_{n_f})^2,n_f]
={\bar a}^{\rm PT}[(Q_{n_f+1}/\Lambda_{n_f+1})^2,n_f+1] \, ,
\end{equation}
which one usually uses in perturbation theory, is modified
and written down as the relation of the contour integrals,
Eq.~(\ref{match-contour}).

As an example, consider a change of the two-loop scale parameter
$\Lambda$ when passing through a quark pair threshold in PT and APT by using
$\alpha(M_{\tau})=0.34 $ and the following values of pole
$c$-, $b$- and $t$-quark masses  $M_c=1.6$~GeV, $M_b=4.5$~GeV, and $M_t=174$~GeV
and $Q_4=2M_c$, $Q_5=2M_b$, $Q_6=2M_t$. In the perturbative
case, we find
$ \Lambda^{\rm PT}_3=419$~MeV,
$ \Lambda^{\rm PT}_4=338$~MeV,
$ \Lambda^{\rm PT}_5=230$~MeV, and
$ \Lambda^{\rm PT}_6=92.6$~MeV,
the ratios of which obey the well-known relations from Ref.~\cite{MC}.
In the APT case, we obtain
$ \Lambda_3^{\rm APT}=607$~MeV,
$ \Lambda_4^{\rm APT}=471$~MeV,
$ \Lambda_5^{\rm APT}=316$~MeV, and
$ \Lambda_6^{\rm APT}=129$~MeV.
The ratios of these quantities are close to the perturbative relations;
however, the PT and the APT values of $\Lambda$ with
the same number of active quarks differ by  about $40\%$.

	       \begin{figure}[hpt]
\centerline{ \psfig{file=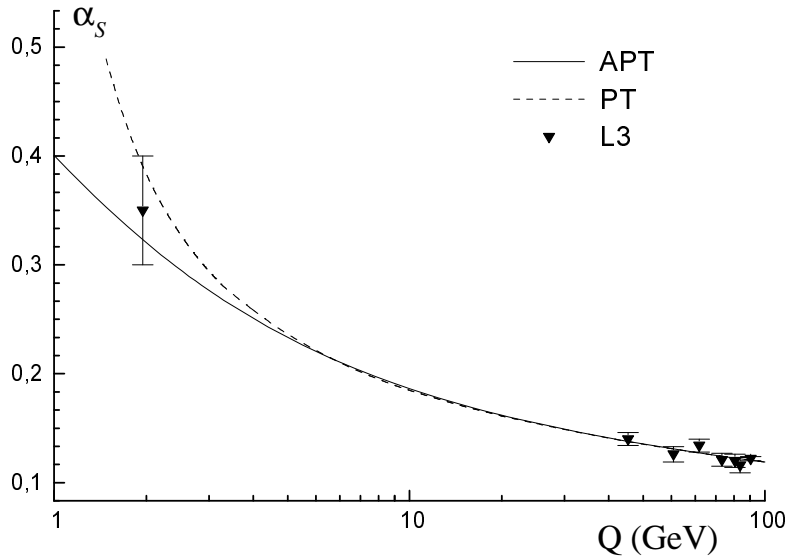,width=13.5cm}}
\caption{\sl QCD evolution of the running coupling constants
(defined in the spacelike region) compared to experimental data.}
       	\label{f7}
\end{figure}

In Fig.~7, we plot results for the QCD evolution of $\alpha_{S}(Q)$,
comparing the APT running coupling constant as discussed above which uses
$s$-channel matching according to Eq.~(\ref{match})
with the standard PT running coupling constant
[see Eq.~(\ref{pt_asimpt})] which uses
the matching procedure given by Eq.~(\ref{pt-match}),
starting from $M_{Z}$ down to $Q=M_{\tau}$.
Also shown on the graph is the experimental data  measured by L3 
collaboration~\cite{L3}.
As experimental input we use the average value
$\alpha_{S}(M_Z)=0.1207 \pm 0.0016$ from Ref.~\cite{L3}.
(In order to do not encumber the figure we do not plot
the corridor of errors). 

We should give a little explanation of how the spacelike APT
running coupling was calculated.  First, we calculate $\Lambda_5$
from the measured value of $\bar\alpha^{\rm APT}(M_Z)=0.1207$.  Then,
we use this value of $\Lambda_5$ to determine $\bar\alpha_s^{\rm APT}$
for all $s$, and hence through the matching procedure, determine
$\Lambda_6$, $\Lambda_4$, $\Lambda_3$.  
The spectral density that we used above at $n_f=3$
 is determined for arbitrary $n_f$ by $\bar a^{\rm APT}_s$ through
$\frac{1}{\displaystyle\pi}\varrho(s,\Lambda_{n_f},n_f)=- s\;
{d \bar a_s(s,\Lambda_{n_f},n_f)}/{d s}$.
An explicit formula for $\varrho$ as a function of $s$, $\Lambda$, and
$n_f$ is given in Eq.~(24) of Ref.\ \cite{MSS1}.
Then, from the spectral representation (\ref{arho}),
we find the spacelike running coupling constant from
\begin{eqnarray}
{\bar \alpha}^{\rm APT}(Q) & =& 4\,\Biggl[
\int_0^{4M_c^2}\,\frac{ds}{s+Q^2}\,\varrho(s,\Lambda_3,3)\, +
\int_{4M_c^2}^{4M_b^2}\,
\frac{ds}{s+Q^2}\,\varrho(s,\Lambda_4,4)\, \\     \nonumber
& & + \int_{4M_b^2}^{4M_t^2}\,\frac{ds}{s+Q^2}\,\varrho(s,\Lambda_5,5)\, +
\int_{4M_t^2}^{\infty}\,
\frac{ds}{s+Q^2}\,\varrho(s,\Lambda_6,6)\,\Biggr] \,.
\end{eqnarray}

It should be stressed that in Fig.~7
we have plotted the value of the QCD running coupling constant extracted
by using the perturbative parametrization. However, this is not really
self-consistent. For instance, in the case of the semileptonic decay of the
$\tau$-lepton, to parametrize the process in the term of the QCD scale
parameter $\Lambda$ one usually uses the analytic properties of the
running coupling, which are obviously broken by the perturbative approximation
due to the unphysical singularities. Within the APT approach it is possible
to maintain the required analytic properties and give a self-consistent
description of the process~\cite{MSS1}. That, in principle, changes
the value of the QCD running coupling extracted from the experimental data.
Thus, the experimental points, plotted in Fig.~7 should be considered
as illustrative only. Nevertheless, it is clear that if we use a
normalization point with a large value of momentum, the curve of the
running coupling constant corresponding to the APT method lies below
than the corresponding PT line, which, from the
 point of view of the perturbative
description, corresponds to a smaller value of $\Lambda$ at low energy.
The fact that low energy data prefer small values of the scale parameter
$\Lambda$ and that at the same time the high energy data prefer larger
values of $\Lambda$ has been emphasized in Ref.~\cite{Shifman}. Thus,
this apparent discrepancy may
be understood in the framework of the APT method; however, we should
mention again that one needs to perform a reanalysis of the low energy
experimental data by using the APT parametrization, as in  the case of
$\tau$ decay~\cite{MSS1}, in order to extract the QCD running coupling
constant.

\section{Conclusions}
Let us briefly summarize our considerations. To determine the
running coupling constant in the timelike region, we took advantage
of APT because it provides a consistent procedure necessary for analytic
continuation. It is to be noted that the APT method ensures not only correct
analytic properties of the running coupling constant but also stability with
respect to higher loop corrections, which is essential for the
stability of our procedure of analytic continuation. This stability
is provided, in part,  by the universal infrared limit value
of the running coupling constant at $q^2=-Q^2\to -0$
that is invariant with respect to higher loop corrections.
The proposed method of constructing the running coupling constant
in the timelike region results in a function with the same universal
infrared limit value when $q^2\to +0$.

Quantitatively, our analysis shows that the effect of analytic
continuation can be associated with $\pi^2$-terms only at very large momentum
transfers of the order of the $Z$-boson mass where the contribution of the
$\pi^2$-terms
is small. At intermediate and, especially, at low momentum transfers
it is important to take account of the correct analytic properties of
$\alpha_{S}$,
which permits a consistent transition into the timelike region.
The $Q^2$-dependence of $\alpha_{S}$
is essentially different from the dependence of $\alpha_S$ in
PT. Our analysis shows that the popular PT expressions for $\alpha_S$
as expansions in $1/{\ln (Q^2/\Lambda^2)}$, containing nonphysical
singularities, do not allow a self-consistent interpretation of
information obtained from different experiments on the evolution of $\alpha_S$
outside of the asymptotic region.
{}From our numerical estimates it follows that analyticity of the
running coupling constant has great influence on the value
of the parameter $\Lambda_{\rm QCD}$ extracted from
experimental data and on the $Q^2$-evolution
of $\alpha_S$. Note that these considerations are also important for the
investigation of power corrections, which are
now under intensive study  (see, e.g., \cite{Grunberg}).
The importance of power corrections in the APT scheme relative to
perturbative terms naturally will be different than in the conventional 
approach. As in PT, the influence of quark thresholds results in a reduction
of the scale parameter $\Lambda$ as the number of active flavors
$n_f$ increases.  However, the importance of maintaining the correct
analytic structure suggests that the required matching be made in
the physical region.

The APT method appears to be fruitful for studying the problem of
analytic continuation of $\alpha_S$ into the timelike region.
There is no doubt that extracting more detailed information from
experimental data on timelike processes requires a more thorough theoretical
analysis within APT including the estimation of the contributions from higher
order processes, mass corrections, and so on. These will be considered in
our subsequent papers.

\section*{Acknowledgements}

We are grateful to D.V. Shirkov, S. V. Mikhailov, R. Ruskov, and I.
L. Solovtsov for useful remarks and interest in the work.
This work has been supported in part by the US National Science
Foundation (grant PHY-9600421) and the US Department of Energy (grant
DE-FG-02-95ER40923).

\end{document}